\documentstyle[twoside]{article}
\catcode`\@=11
\long\def\@makefntext#1{
\protect\noindent \hbox to 3.2pt {\hskip-.9pt
$^{{\eightrm\@thefnmark}}$\hfil}#1\hfill}               

\def\@makefnmark{\hbox to 0pt{$^{\@thefnmark}$\hss}}    

\def\ps@myheadings{\let\@mkboth\@gobbletwo
\def\@oddhead{\hbox{}
\rightmark\hfil\eightrm\thepage}
\def\@oddfoot{}\def\@evenhead{\eightrm\thepage\hfil
\leftmark\hbox{}}\def\@evenfoot{}
\def\sectionmark##1{}\def\subsectionmark##1{}}

\oddsidemargin=\evensidemargin
\newcounter{sectionc}\newcounter{subsectionc}\newcounter{subsubsectionc}
\renewcommand{\section}[1] {\vspace{12pt}\addtocounter{sectionc}{1}
\setcounter{subsectionc}{0}\setcounter{subsubsectionc}{0}\noindent
        {\tenbf\thesectionc. #1}\par\vspace{5pt}}
\renewcommand{\subsection}[1] {\vspace{12pt}\addtocounter{subsectionc}{1}
        \setcounter{subsubsectionc}{0}\noindent
        {\bf\thesectionc.\thesubsectionc. {\kern1pt \bfit #1}}\par\vspace{5pt}}
\renewcommand{\subsubsection}[1] {\vspace{12pt}\addtocounter{subsubsectionc}{1}
        \noindent{\tenrm\thesectionc.\thesubsectionc.\thesubsubsectionc.
        {\kern1pt \tenit #1}}\par\vspace{5pt}}
\newcommand{\nonumsection}[1] {\vspace{12pt}\noindent{\tenbf #1}
        \par\vspace{5pt}}

\newcounter{appendixc}
\newcounter{subappendixc}[appendixc]
\newcounter{subsubappendixc}[subappendixc]
\renewcommand{\thesubappendixc}{\Alph{appendixc}.\arabic{subappendixc}}
\renewcommand{\thesubsubappendixc}
        {\Alph{appendixc}.\arabic{subappendixc}.\arabic{subsubappendixc}}

\renewcommand{\appendix}[1] {\vspace{12pt}
        \refstepcounter{appendixc}
        \setcounter{figure}{0}
        \setcounter{table}{0}
        \setcounter{lemma}{0}
        \setcounter{theorem}{0}
        \setcounter{corollary}{0}
        \setcounter{definition}{0}
        \setcounter{equation}{0}
        \renewcommand{\thefigure}{\Alph{appendixc}.\arabic{figure}}
        \renewcommand{\thetable}{\Alph{appendixc}.\arabic{table}}
        \renewcommand{\theappendixc}{\Alph{appendixc}}
        \renewcommand{\thelemma}{\Alph{appendixc}.\arabic{lemma}}
        \renewcommand{\thetheorem}{\Alph{appendixc}.\arabic{theorem}}
        \renewcommand{\thedefinition}{\Alph{appendixc}.\arabic{definition}}
        \renewcommand{\thecorollary}{\Alph{appendixc}.\arabic{corollary}}
        \renewcommand{\theequation}{\Alph{appendixc}.\arabic{equation}}
        \noindent{\tenbf Appendix \theappendixc #1}\par\vspace{5pt}}
\newcommand{\subappendix}[1] {\vspace{12pt}
        \refstepcounter{subappendixc}
        \noindent{\bf Appendix \thesubappendixc. {\kern1pt \bfit #1}}
        \par\vspace{5pt}}
\newcommand{\subsubappendix}[1] {\vspace{12pt}
        \refstepcounter{subsubappendixc}
        \noindent{\rm Appendix \thesubsubappendixc. {\kern1pt \tenit #1}}
        \par\vspace{5pt}}

\newcommand{\textlineskip}{\baselineskip=13pt}
\newcommand{\smalllineskip}{\baselineskip=10pt}

\def\eightcirc{
\begin{picture}(0,0)
\put(4.4,1.8){\circle{6.5}}
\end{picture}}
\def\eightcopyright{\eightcirc\kern2.7pt\hbox{\eightrm c}}

\newcommand{\copyrightheading}[1]
        {\vspace*{-2.5cm}\smalllineskip{\flushleft
        {\footnotesize Modern Physics Letters A, LA-UR-98-2905 and EFUAZ FT-98-62}\\
        {\footnotesize $\eightcopyright$\, World Scientific Publishing
         Company}\\
         }}

\def\abstracts#1#2#3{{
        \centering{\begin{minipage}{4.5in}\baselineskip=10pt\footnotesize
        \parindent=0pt #1\par
        \parindent=15pt #2\par
        \parindent=15pt #3
        \end{minipage}}\par}}

\newcommand{\bibit}{\nineit}
\newcommand{\bibbf}{\ninebf}
\renewenvironment{thebibliography}[1]
        {\frenchspacing
         \ninerm\baselineskip=11pt
         \begin{list}{\arabic{enumi}.}
        {\usecounter{enumi}\setlength{\parsep}{0pt}
         \setlength{\leftmargin 12.7pt}{\rightmargin 0pt} 
         \setlength{\itemsep}{0pt} \settowidth
        {\labelwidth}{#1.}\sloppy}}{\end{list}}

\newcounter{itemlistc}
\newcounter{romanlistc}
\newcounter{alphlistc}
\newcounter{arabiclistc}

\newcommand{\fcaption}[1]{
        \refstepcounter{figure}
        \setbox\@tempboxa = \hbox{\footnotesize Fig.~\thefigure. #1}
        \ifdim \wd\@tempboxa > 5in
           {\begin{center}
        \parbox{5in}{\footnotesize\smalllineskip Fig.~\thefigure. #1}
            \end{center}}
        \else
             {\begin{center}
             {\footnotesize Fig.~\thefigure. #1}
              \end{center}}
        \fi}

\newcommand{\tcaption}[1]{
        \refstepcounter{table}
        \setbox\@tempboxa = \hbox{\footnotesize Table~\thetable. #1}
        \ifdim \wd\@tempboxa > 5in
           {\begin{center}
        \parbox{5in}{\footnotesize\smalllineskip Table~\thetable. #1}
            \end{center}}
        \else
             {\begin{center}
             {\footnotesize Table~\thetable. #1}
              \end{center}}
        \fi}

\def\@citex[#1]#2{\if@filesw\immediate\write\@auxout
        {\string\citation{#2}}\fi
\def\@citea{}\@cite{\@for\@citeb:=#2\do
        {\@citea\def\@citea{,}\@ifundefined
        {b@\@citeb}{{\bf ?}\@warning
        {Citation `\@citeb' on page \thepage \space undefined}}
        {\csname b@\@citeb\endcsname}}}{#1}}

\newif\if@cghi
\def\cite{\@cghitrue\@ifnextchar [{\@tempswatrue
        \@citex}{\@tempswafalse\@citex[]}}
\def\citelow{\@cghifalse\@ifnextchar [{\@tempswatrue
        \@citex}{\@tempswafalse\@citex[]}}
\def\@cite#1#2{{$\null^{#1}$\if@tempswa\typeout
        {IJCGA warning: optional citation argument
        ignored: `#2'} \fi}}

\def\@refcitex[#1]#2{\if@filesw\immediate\write\@auxout
        {\string\citation{#2}}\fi
\def\@citea{}\@refcite{\@for\@citeb:=#2\do
        {\@citea\def\@citea{, }\@ifundefined
        {b@\@citeb}{{\bf ?}\@warning
        {Citation `\@citeb' on page \thepage \space undefined}}
        \hbox{\csname b@\@citeb\endcsname}}}{#1}}

\def\@refcite#1#2{{#1\if@tempswa\typeout
        {IJCGA warning: optional citation argument
        ignored: `#2'} \fi}}

\def\refcite{\@ifnextchar[{\@tempswatrue
        \@refcitex}{\@tempswafalse\@refcitex[]}}

\def\pmb#1{\setbox0=\hbox{#1}
        \kern-.025em\copy0\kern-\wd0
        \kern.05em\copy0\kern-\wd0
        \kern-.025em\raise.0433em\box0}

\def\fnm#1{$^{\mbox{\scriptsize #1}}$}
\def\fnt#1#2{\footnotetext{\kern-.3em
        {$^{\mbox{\scriptsize #1}}$}{#2}}}

\def\fpage#1{\begingroup
\voffset=.3in
\thispagestyle{empty}\begin{table}[b]\centerline{\footnotesize #1}
        \end{table}\endgroup}

\def\runninghead#1#2{\pagestyle{myheadings}
\markboth{{\protect\footnotesize\it{\quad #1}}\hfill}
{\hfill{\protect\footnotesize\it{#2\quad}}}}
\font\tenrm=cmr10
\font\tenit=cmti10
\font\tenbf=cmbx10
\font\bfit=cmbxti10 at 10pt
\font\ninerm=cmr9
\font\nineit=cmti9
\font\ninebf=cmbx9
\font\eightrm=cmr8

\def\qed{\hbox{${\vcenter{\vbox{                        
   \hrule height 0.4pt\hbox{\vrule width 0.4pt height 6pt
   \kern5pt\vrule width 0.4pt}\hrule height 0.4pt}}}$}}

\begin{document}

\def\be{\begin{equation}}
\def\ee{\end{equation}}


\runninghead{D. V. Ahluwalia}
{Non-locality and gravity-induced CP violation}
\normalsize\textlineskip
\thispagestyle{empty}
\setcounter{page}{1}

\copyrightheading{}                     

\vspace*{0.88truein}

\fpage{1}

\centerline{\bf NON-LOCALITY  AND GRAVITY-INDUCED CP VIOLATION}
\vspace*{0.37truein}

\centerline{\footnotesize D. V. AHLUWALIA\fnm{*}\fnt{*}
{E-mail: ahluwalia@phases.reduaz.mx, 
vahluwa@cantera.reduaz.mx,  av@p25hp.lanl.gov}
}
\vspace*{0.015truein}
\baselineskip=10pt
\centerline{\footnotesize\it
Escuela de Fisica, Univ. Aut. de Zacatecas, Apartado Postal C-580}
\vspace*{0.015truein}
\centerline{\footnotesize\it
Zacatecas, ZAC 98068, Mexico}

\vspace*{0.015truein}

\centerline{\footnotesize\it and}

\vspace*{0.015truein}
\baselineskip=10pt
\centerline{\footnotesize\it
Physics Division P-25, Mail Stop H-846}

\vspace*{0.015truein}
\centerline{\footnotesize\it Los Alamos National Laboratory,
New Mexico 87545, USA}


\vspace*{0.21truein}

\def\po{\mathaccent 23 p^\mu}
\def\e{\mbox{e}}

\abstracts{
We consider a space-time dependent relative phase  between the
right- and left-handed spinors and show that it results in a
violation of locality in the presence of gravity
once the demand of parity covariance is dropped.
This violation of locality is such that it readily
interprets itself as a gravity-induced CP violation, and at the
same time confirms an earlier remark by Wigner that a
representation space carries more information than a wave equation.
This happens, as Kirchbach has noted, because while the 
dimensionality of an irreducible representation space does not depend
upon the concrete realization of the symmetry generators, Noether currents
(Dirac, versus Majorana, versus the CP violating construct presented here) do.
The gravity-induced CP violation  provides a dynamical reason
on how a neutron star carrying its baryon and lepton numbers can
collapse into a black hole and loose information on the latter
characteristics.}{}{}

\section{Introduction and historical backgound}

The uncertainty relation $\Delta x\,
\Delta p_x \sim \hbar$ is  a direct
consequence of the fundamental commutator $\left[ x,\,p_x\right]=i
\hbar$ between the position and  momentum of a particle.
Its role in the foundations of physics can hardly be
overemphasized. Yet, as the century that began with this profound
change in our understanding of Nature reaches towards its end, we
are beginning to realize hints for further deeper changes. It is
becoming increasingly clear that in the presence of gravitation the
fundamental uncertainty relations, and hence the fundamental
commutators, must be modified [\refcite{mod1}-\refcite{mod14}]. For
instance, 
in the context of string theories 
$\Delta x\, \Delta p_x
\sim \hbar$ gets replaced by
\begin{equation}\Delta x\,
\Delta p_x \sim \hbar\left[1 +\left(\lambda_P
\,\Delta p_x/\hbar\right)^2\right]
,\label{witten}
\end{equation}
with $\lambda_P$ of the order of the Planck length $\sqrt{\hbar
G/c^3}$. However,  as emphasised by Witten\cite{mod14} a proper
theoretical
framework for the extra term in the uncertainty relation has not
yet emerged.

Once modifications of the uncertainty relations are taken
seriously, the question naturally arises if such can also occur within
the context of the existing point-particle quantum field theoretic
framework, in four-dimensional space-time, with minimal changes.
Modifications to the uncertainty relations of the type considered
in (\ref{witten}) clearly imply that the canonical bosonic
commutators, and the fermionic anticommutators, in being
mathematical expressions of the {\it locality} of the underlying
quantum field theory,\cite{Schwinger} have to change to
incorporate non-locality.

Our answer to the question posed in the opening remark of the
preceding paragraph is a well-formulated and clear `yes.' The
fundamental modifications to the structure of
quantum field theory can be induced by gravity
and occur via a non-locality that resides in a
certain space-time derived fields. The origin of the presented
modifications goes beyond simply considering a given
flat-space-time field in a curved background. The precise nature of
these statements shall become clear in due course.

We here show that an allowed relative phase between the right- and
left-handed spinors that may appear in covering the
full Lorentz group without demanding parity covariance,
can serve as a source for non--locality in the presence of gravity.
The violation of locality produced in this way is such that it readily
interprets itself as a gravity-induced CP violation, and at the
same time confirms an earlier remark by Wigner that a
representation space carries more information than a wave 
equation.\cite{EPW1963}

A few historical remarks are now in order. Even before the
experimental discovery of CP violation,\cite{CPex} the
possibility was considered by Good that gravitation may induce CP
violation\cite{Good} and he used the then-existing circumstances
to  study  the gravitational behavior of 
antimatter.
Since then the idea of gravity-induced
CP violation has been repeatedly considered 
[\refcite{repeat1}-\refcite{repeat5}]. 
The Good framework is built
upon the Morrison-Gold conjecture presented in their celebrated
1957 Gravity Research Foundation essay that particles and
antiparticles may carry opposite gravitational masses.\cite{MG} 
Note that the appearance of different
particle--antiparticle gravitational masses within the
gravity-induced CP violation scenario of Good suggests violation of
the principle of equivalence. In a  recent
work Chardin  argued that gravity-induced CP violation
provides a parameter free explanation of the CP violation observed
in the neutral Kaon system.\cite{Chardin} 
In the Chardin argument no violation of the equivalence principle 
occurs. Instead, gravitational repulsion naturally emerges in the
context of wormholes and the Kerr geometry.
Indpendently, Fischbach
{\it et al.} have arrived at a similar conclusion.\cite{FT}

A further argument in favor of gravitation as a source of CP
violation can be found in Hawking's work where  black holes are
shown to posses a thermodynamic entropy;\cite{SH} and
consequently lead to  T- (and CP-, in a CPT preserving framework)
violating processes.

In this paper we show how to drop the Morrison-Gold conjecture in a
non--trivial way and obtain a gravity-induced CP violation  within
a CPT covariant framework. Since the principle of equivalence is no
longer violated within the present scheme, the proposed
gravity-induced CP violation is fundamentally different from the
Morrison-Gold framework. Since no reference is made to wormholes
and Kerr geometry, the presented framework has greater generality than
the Chardin-Fischbach scheme.

To keep the physics transparent our thesis  shall be presented in
flat space-time, and  gravitation  considered in the weak field
limit {\it a la} Sakurai (see [\refcite{JJS}], pp. 126-129), on the one
side, and in the spirit of experimental work on gravitationally
induced quantum interference,\cite{W} on the other side. A
further reason for this approach resides in the fact that {\sl if
one confines oneself to a purely general relativistic framework},
certain physically observable quantum mechanical phases can become
non--observable. In essence, this means that the general
relativistic description of gravitation may not be considered
complete in the quantum realm\cite{essential,EPRg} --- see section 2.4
for further details.

\nobreak

\section{A space-time origin of non-locality}
\nobreak

The thesis that an origin of CP violation is to be found at the
level of the representations of the Lorentz group, and  that in the
presence of gravity this violation is deeply connected with the
space-time metric shall be made in three parts. In the first part
an unsuspected space-time dependent relative phase between the
right- and left-handed spinors is introduced. In the second part
it is discovered that this phase is deeply connected with the C, P, and
T properties of the $(1/2,0)\oplus(0,1/2)$ representation space, and that
despite CPT-covariance it carries in it an essential element of
non-locality. In the third part a physical interpretation is put
forward in which the space-time dependent relative phase induces
deviations of the metric tensor from its flat space-time form.
As a result a gravity-induced CP violating
structure emerges which has  profound consequences for astrophysical and
cosmological processes.\fnm{a}\fnt{a}{
In this, but not the following, section we shall set $\hbar$ and
$c$ equal unity.} This thesis spans sections 2.1 to 2.3. Section 
2.4 then provides a parenthetic argument regarding the observability
of constant gravitational potentials in the context of an earlier
work of Kenyon.\cite{repeat4}

\subsection{New Relative phases between the right- and left-handed spinors}

\noindent
To present our thesis let us begin by recalling  that in the recent
generalization of the Case-McLennan reformulation   ( see  Refs. 
[\refcite{EMa}-\refcite{VVD}])
 of the Majorana field the
  {it
relative phase} (to be denoted by $\zeta$ here) between the right- and
left-handed spinors plays an important physical role:\fnm{b}
\fnt{b}{In Eq. (\ref{os}), $\Theta_{[j]}$ is the Wigner time-reversal
operator,
$
\Theta_{[j]}\,{\vec J}\,\Theta_{[j]}^{-1}\,=\,-\,{\vec J}^\ast
$. It is a consequence of this property that if
$\phi_{_{L}}(p^\mu)$ transforms as a $(0,\,j)$ co--spinor under
Lorentz boosts, the construct
$\Theta_{[j]}\,\phi_{_{L}}^\ast(p^\mu)$ transforms as a $(j,\,0)$
spinor, i.e. as a right-handed spinor. Similarly,  if
$\phi_{_{R}}(p^\mu)$ transforms as a $(j,\,0)$ spinor, then the
construct $\Theta_{[j]}\,\phi_{_{R}}^\ast(p^\mu)$ transforms as
a $(0,\,j)$ co-spinor, i.e., as a left-handed  spinor. The notation
$\vec{J}$ stands for
the usual $(2j+1)\times (2j+1)$ spin-$j$ matrices.}
\begin{equation}
\lambda(p^\mu)\,\equiv
\left(
\begin{array}{c}
\left(\zeta_\lambda\,\Theta_{[j]}\right)\,\phi^\ast_{_L}(p^\mu)\\
\phi_{_L}(p^\mu)\\
\end{array}
\right)\,\,,\quad
\rho(p^\mu)\,\equiv
\left(
\begin{array}{c}
\phi_{_R}(p^\mu)\\
\left(\zeta_\rho\,\Theta_{[j]}\right)^\ast\,\phi^\ast_{_R}(p^\mu)
\end{array}
\right)
\,\,\quad .\label{os}
\end{equation}
Indeed, for fermion fields these phases must take on the values
$\pm\, i$ to ensure that the spinors of the $(j,0)\oplus(0,j)$
representation are self/anti-self charge conjugate, i.e., they are
of the extended Majorana type. It is this phase, equal to $\pm
\,i$, that emerges as the intrinsic parity of the Majorana
particles in the original analysis of Racah.\cite{GR} On the
other hand, for Dirac spinors,
 within the wisdom of the 1960s\cite{LHR1992} it was
argued by Ryder in his  recent textbook on quantum field theory
that (see Ref. [\refcite{LHR1987}], p. 44),
\begin{quote}
{\it Now when a particle is at rest, one cannot define its spin as
either left- or right-handed, so
\begin{equation}\phi_R(\vec{0})=
\phi_L( \vec{0}).\label{rb}\end{equation}}
\end{quote}
Here $\vec 0$ represents the vanishing three momentum of the
particle at rest, while its four momentum is represented by $\po
\equiv
\{m,\,\vec{0}\}$.

We now make the crucial observation of this paper. While the above
quoted argument remains valid in the classical domain, it does not
contain the full physics allowed by the relativistic
quantum-mechanical framework.\cite{LHR_R} 
In the quantum realm it must be
generalized to read:
\begin{quote}
{\it
 The full exploitation of the relativistic quantum-mechanical
framework allows the equality expressed by Eq. (\ref{rb}) up to a
phase,}
\begin{equation}\bf
\phi_R(\po)=\pm \exp\left[\pm\, i\,\phi(x)\right]
 \phi_L(\po) .\label{phasea}
\end{equation}

\end{quote}
\noindent
The phase, $\phi(x)$, may have a  space-time dependence.
While the  existence of this phase is permitted by the quantum
mechanical framework, its space-time dependence is a requirement of
relativity
-- the arguments for this dual requirement are similar to the
standard textbook arguments that are found for the introduction of
space-time dependent ``local'' gauge transformation 
(see Ref. [\refcite{LHR1987}], Sec. 3.3). 
The first $\pm$ sign on the right-hand side of
Eq. (\ref{phasea}) is introduced to span the full $(j,0)\oplus(0,j)$
representation space. In the second $\pm$ sign of Eq. (\ref{phasea}),
the plus sign corresponds to the particles (the `positive energy
solution') and the negative sign to the anti-particles (the
`negative energy solution') -- see below.

It is this observation, we shall argue, that contains the
answer-in-affirmative to the question posed in Sec. 1 above. In its
physical essence, this observation parallels the ideas of Weyl,\cite{HW} and
Yang and Mills.\cite{YM}
What follows is a mathematical  exercise  to distill the physical
content of the generalization contained in Eq.
(\ref{phasea}).\fnm{c}\fnt{c}{Equation (\ref{phasea}) 
owes its origins to
discussions with C. Burgard when we both were at Texas A\&M
University in the early nineties. It is my pleasure to have this
opportunity to acknowledge this with thanks.}

If the phase $\phi(x)$ has to carry a physical meaning, its space-time
dependence must be associated with some dynamical property. The
phase $\phi(x)$ may be interpreted either as a Higgs-like
field\cite{RE}, or, as done in Sec. 2.3 below, as the
deviation the metric tensor in the weak--filed limit
from its flat space-time form
$\eta_{\mu\nu}=\mbox{diag.}\left(1,-1,-1,-1 \right)$.
Whether the Higgs field is some manifestation of gravitation in disguise
remains a pregnant possibility.

The right- and left-handed spinors, without  reference to a wave
equation or a Lagrangian, Lorentz transform 
as:\cite{LHR1987,dva_thesis}
\begin{equation}
 \phi_{R}(p^\mu) = \exp\left(+\vec{J}\cdot{\vec{\varphi}}\right)
                   \,\phi_{ R}(\po)\,,\quad
 \phi_{ L}(p^\mu) = \exp\left(-\,\vec{J}\cdot{\vec{\varphi}}\right)
\,\phi_{ L}(\po).\label{zb}
\end{equation}
Here, $p^\mu$ represents the boosted energy-momentum vector
$(E,\vec{p}\,)$. The boost parameter, $\vec\varphi$, is defined
as
\begin{equation}
\cosh(\varphi)=\gamma=(1-v^2)^{- {1/2}} ={E\over m},\quad
\sinh(\varphi)=v \gamma= {{\vert{\vec{p}}\,\vert}\over m},
\end{equation}
with $\widehat{\varphi}={\vec{p}/{\vert\vec{p}\,\vert}}$, and
$\vec{v}$ stands for  the velocity of the particle.

The considerations so far are generally true for any spin. To
present our thesis it suffices to confine to spin
one-half (however, note that an
extension to higher spins may contain important physics).

\nobreak
\subsection{Non-locality for spin ${\bf 1/2}$}

\noindent
For spin one-half,  when Eqs. (\ref{phasea}) and (\ref{zb}) are coupled
{\it a la} Ryder (see Ref. [\refcite{LHR1987}], Sec. 2.3),
 we obtain on setting $\vec{J}=\vec{\sigma}/2$,
 \nobreak
\begin{eqnarray}
-\,\zeta^{-1}(x) \,\phi_{ R}\left(p^\mu\right) + \left(
        {{E+\vec{\sigma}\cdot \vec{p}}\over{m}}\right)
\phi_{ L}\left(p^\mu\right) &&=0,\label{c1}\\
\left({{E-\vec{\sigma}\cdot \vec{p}}\over{m}}\right)
\phi_{ R}\left(p^\mu\right)
-\zeta(x)\, \phi_{ L}\left(p^\mu\right) &&=0.\label{c2}
\end{eqnarray}
In Eqs. (\ref{c1}) and (\ref{c2}), $\zeta(x)$ stands for the phase factor 
$\pm \exp\left[\pm\, i\,\phi(x)\right]$. Now we introduce the
$(1/2,0)\oplus(0,1/2)$ Weyl-representation  spinor 
[{\it cf.} Weyl-representation
extended Majorana spinors of Eq. (\ref{os})],
\begin{equation}
\psi\left(p^\mu\right) =
\left(
\begin{array}{c}
\phi_{ R}\left(p^\mu\right) \\
\phi_{ L}\left(p^\mu\right)
\end{array}
\right),\label{nds}
\end{equation}
{\it with $\phi_R(\po)$ and $\phi_L(\po)$ related by Eq. (\ref{phasea}).}
Thus, note is to be immediately taken that in case of the arbitrary
phases, $\phi(x)$, the spinors (\ref{nds}), are {\it not} identical
to the standard Dirac spinors. By the end of this paper the reader
shall infer that these spinors become Dirac spinors when $\phi(x)$
acquires a space-time independent value that is an integral
multiple of $2\pi$.

Eqs.(\ref{c1}) and (\ref{c2}) can be cast into the compact form,
\begin{equation}
\left(\gamma^\mu\,p_\mu\,-\,\xi(x)\,m\right)
\psi\left(p^\mu\right) \,=\,0\quad\, , \label{d}
\end{equation}
where $\gamma^\mu$ are the Dirac's gamma matrices in the Weyl
representation (see Ref. [\refcite{LHR1987}], Eq. 2.92),
and $\xi(x)$ is a $4\times 4$ matrix,
\begin{equation}
\xi(x) = \left(\begin{array}{cc}
\zeta^{-1}(x)\,I_2 & 0_2\\
0_2 & \zeta(x)\,I_2
\end{array}\right)\,
.\label{f}
\end{equation}
We use the notation $I_n$ for the $n\times n$ unit
matrix, while $0_n$ stands for a $n\times n$ zero- matrix.

\vspace*{0.21truein}
\noindent
Several observations are immediately in order.

(a) Equation (\ref{d}) has been derived from the most basic
space-time arguments. In particular, no Lagrangian needed to be
postulated.. A Lagrangian may be constructed for field theoretic
purposes  which yields Eq. (\ref{d}). The hermiticity of this
Lagrangian is assured because, apart from the known properties of
$\gamma^\mu$, we also have $\gamma^0 \xi(x)=
\xi^\dagger(x)\gamma^{0\,\dagger}$.

(b)
While Eq. (\ref{d}) is CPT covariant, it is not P covariant for an
arbitrary  phase $\phi(x)$. The parity covariance
requires
\begin{equation}
\gamma^0\,\xi(x)\,\gamma^0=\xi(x^\prime),
\end{equation}
where $x^\prime$ is the parity transformed $x$. Tentatively assuming
$\zeta(x)$ to be an even function of $x$ 
(without {\it a priori} justification),
this requirement reduces to the condition $\zeta^{-1}(x)
=\zeta(x)$, and implies  $\zeta
=\pm \,1$. In terms of the phase, this translates to
a space-time independent $\phi=0$, or more generally an integral
multiple of $2\pi$ --
 {\it cf.} Ref. [\refcite{RE}]. Further, the phase $\pm\,i$ encountered in
the case of the extended Majorana spinors\cite{A} is an exact
counterpart of the $\pm
\,1$ for the Dirac spinors. For Dirac spinors it refers to the
relative parities of fermions and anti-fermions. For the extended
Majorana spinors this phase is the parity of the self/anti-self
charge conjugate particles. The relative phase of $\pm i$ appearing
for Majorana fields is determined by the requirement of
self/anti-self conjugacy under the operation of charge conjugation,
$C$. For the Dirac fields the phase of $\pm\,1$ follows from the
requirement that the respective spinors be eigenstates of the
charge operator, $Q$. The physical origin for the different
intrinsic parities of the Dirac particle and anti--particle, as
well as of the extended Majorana spinors, arises from the fact that
the latter are eigenstates of the charge conjugation operator,
while the former are eigenstates of the charge operator -- and that
$C$ and $Q$ do not commute, $\left[ C,\,Q\right]\ne 0$. The
requirement of parity covariance collapses the phase field
$\zeta(x)$ to the constant value of $\,\pm 1$ throughout the
space-time.
 This means that the {\it
measurement} of parity collapses the relative phase factor between the
right- and left-handed spinors to the constant eigenphases $\pm
1$. Here is the first hint that the demand of parity covariance 
is related to the locality. If the demand for parity covariance is
dropped the resulting theory may become non-local
 (with the understanding that this
non-locality is of a similar origin as that involved in the 
``collapse of a wave function''). In particular,

\begin{equation}
\psi\left[p^\mu, \phi(x)\rightarrow 0,2\pi,\cdots\right]
\rightarrow \psi(p^\mu)_{Dirac} .
\end{equation}

(c) We find
\begin{equation}
\mbox{Det\,}
|\gamma^\mu\,p_\mu\,-\,\xi\,m|
=\left({\vec{p}}^{\,2} +m^2-E^2\right)^2.
\end{equation}
The independence of $\mbox{Det\,}|\gamma^\mu\,p_\mu\,-\,\xi\,m | $
on the phase $\phi(x)$ enables one to give on the classical level a
physical interpretation of the CP-violating fields in terms of
particles and antiparticles, i.e., in terms of positive- and
negative-energy solutions of Eq. (\ref{d})  with
$E=\pm\sqrt{{\vec{p}\,}^{\,2} + m^2}$.

(d) Because
\begin{equation}
\exp\left[i\gamma^5\phi(x)\right]
\left(
\begin{array}{c}
\phi_{R}\left(p^\mu\right) \\
\phi_{L}\left(p^\mu\right)
\end{array}
\right) = \exp\left[i\phi(x)\right]
\left(
\begin{array}{c}
\phi_{R}\left(p^\mu\right) \\
\exp\left[-i 2 \phi(x)\right]\phi_{ L}\left(p^\mu\right)
\end{array}
\right)\label{chiral}
\end{equation}
the chiral transformation introduces a relative phase between the
right- and left-handed spinors. However, it is important to note
that Eq.~(\ref{d}) differs from the chirally transformed Dirac
equation.

\vspace*{0.21truein}

The $\zeta(x)$ for the ``particle'' $u$-spinors is, $\zeta_u(x) =
+\,\exp\left[+\,i\,\phi(x)\right]$, and for the antiparticle
$v$-spinors it is, $\zeta_v(x)= -\,\exp\left[-\,i\,\phi(x)\right]$.
We thus have (in the usual notation), using Eqs. (\ref{phasea}),
(\ref{zb}), and (\ref{nds}):
\begin{eqnarray}
u_{+ 1/2}(p^\mu) = A \left(
\begin{array}{c}
E_+\\ p_+\\ E_-\,\e^{-\,i\,\phi(x)}\\
-\,p_+\,\e^{-\,i\,\phi(x)}
\end{array}
\right),
u_{- 1/2}(p^\mu) = A\left(
\begin{array}{c}
p_-\\ E_-\\
-\,p_-\,\e^{-\,i\,\phi(x)}\\
E_+\,\e^{-\,i\,\phi(x)}
\end{array}
\right),\\
v_{+ 1/2}(p^\mu) = A
\left(
\begin{array}{c}
E_+\\ p_+\\
-\,E_-\,\e^{+\,i\,\phi(x)}\\
p_+\,\e^{+\,i\,\phi(x)}
\end{array}
\right),
v_{- 1/2}(p^\mu) = A\left(
\begin{array}{c}
p_-\\ E_-\\ p_-\,\e^{+\, i\,\phi(x)}\\
-\,E_+\,\e^{+\, i\,\phi(x)}
\end{array}
\right),
\end{eqnarray}
with $A=\left[2(m+E)\cos(\phi(x))\right]^{-1/2}$, $E_\pm=E+m\pm
p_z$, and $p_\pm=p_x\pm i\, p_y$. These spinors have the standard
norm:
\begin{equation}
\overline{u}_\sigma(p^\mu)\,u_\sigma(p^\mu) =
+\,2\, m\, \delta_{\sigma\sigma^{\prime}}\,\,
\mbox{and}\,\,
\overline{v}_\sigma(p^\mu)\,v_\sigma(p^\mu) =
-\,2\, m\, \delta_{\sigma\sigma^{\prime}}.
\end{equation}
As usual,  $\overline{\psi}(p^\mu)=\psi^\dagger(p^\mu)\,\gamma^0$.

It is to be explicitly noted that $\psi(p^\mu)$ contains an
implicit $x$ dependence via Eq. (\ref{phasea}). This fact requires extra
care in obtaining configuration-space representation. However, in
the remainder  of this essay we are only concerned in obtaining
consequences for an experimental region over which $\phi(x)$ is
essentially constant. Under these circumstances Eq. (\ref{d})
yields the ``CP violating Dirac equation'' postulated by Funakubo {\it
et al.}  in the cosmological context.\cite{cp}

Now to study the structure of the theory as regards locality, the
field operator has to be constructed first:\fnm{d}\fnt{d}{
In order to account for the translational invariance of the
arguments
-- see Ref. [\refcite{SWQFT}], the algebra of the Lorentz group
has to be extended to incorporate the generators of the space-time
translations. In writing $\Psi(x)$
we assume, for the moment and as already indicated,
that we are only interested
in application to an experiment confined to a region of space-time
over which $\phi(x)$ is essentially constant.}

\begin{eqnarray}
\Psi(x)&=&
 \int\frac{\mbox{d}^3 k}{(2\pi)^3} {{m}\over{k_0}}
\sum_{\sigma = +1/2,-1/2}\nonumber\\
&\times& \left[ b_\sigma(k^\mu)\, u_\sigma(k^\mu)
\,\exp(-i k_\mu x^\mu)+ d^\dagger_\sigma(k^\mu)\, v_\sigma(k^\mu)
\,\exp(+i k_\mu x^\mu)\right].\end{eqnarray}
\nobreak
For the {\it particle interpretation} of the theory, the following
equal-time anticommutators are assumed to be satisfied,
\begin{equation}
\left\{b_\sigma(k^\mu),\,b^\dagger_{\sigma^\prime}(k^\mu)\right\}
=
\left\{d_\sigma(k^\mu),\,d^\dagger_{\sigma^\prime}(k^\mu)\right\}
=(2\,\pi)^3 \frac{k_0}{m} \delta^3(\vec{k} -\vec{k}^\prime)
\delta_{\sigma\sigma^\prime}\,,\label{ac1}\\
\end{equation}
with the remaining four anticommutators being zero. With these
definitions the {\it non-locality} of the theory becomes manifest
on calculating the anticommutator $\left\{\Psi_i(\vec{
x},\,t),\,\Psi^\dagger_j (\vec x^{\,\prime},\, t)\right\}$.
Exploiting the easily derivable
 identities,
\begin{eqnarray}
&&\sum_\sigma u_\sigma(p^\mu)\,\overline{u}_\sigma(p^\mu) =
{1\over{\cos\left(\phi(x)\right)}}
\left(\gamma_\mu p^\mu +\xi^{-1}_u(x)\,m\right),\\
&&\sum_\sigma v_\sigma(p^\mu)\,\overline{v}_\sigma(p^\mu) =
\frac{1}{\cos\left(\phi(x)\right)}
\left(\gamma_\mu p^\mu +\xi^{-1}_v(x)\,m\right),
\end{eqnarray}
and exploiting the standard textbook techniques (see, e.g., Ref.
[\refcite{LHR1987}], Sec. 4.3), one is led to
\begin{equation}
\left\{\Psi_i(\vec{x},\,t),\,\Psi^\dagger_j
(\vec{x}^{\,\prime},\, t)\right\} = \frac{2
m}{\cos\left(\phi(x)\right)}
\left(
\delta_{ij} \,+\, {\cal O}_{ij}
\right)
\delta^3\left(\vec{x}-\vec{x}^{\,\prime}\right).\label{nl}
\end{equation}
In Eq. (\ref{nl}), ${\cal O}_{ij}$ is purely {\it off}-diagonal,
\begin{equation}
{\cal O}_{ij} \equiv
\frac{i\, m}{k_0}
\left(
\begin{array}{cc}
0_2 & I_2 \\
-\,I_2 & 0_2
\end{array}
\right)\,\sin\left(\phi(x)\right)
.\label{acpsi}
\end{equation}
The non-locality completely resides in ${\cal O}_{ij}$. It vanishes
when (a) the phase $\phi(x)$ is an exact integral multiple of $\pi$, and/or
(b) the particle is massless, or $m/k_0$ approaches zero. Further,
the non-locality manifests in the spinorial space (i.e., the
spinorial indices) and not in the configuration space (i.e, the
$\vec x$ space).\fnm{e}\fnt{e}{It is to be parenthetically  noted
that the particle interpretation contained in Eqs. (\ref{ac1}) may
breakdown for $\sin(\phi(x))\sim 1$ in which case Eqs. (\ref{ac1})
would need to contain a non-trivial $\phi(x)$ dependence. This
observation is dictated by the fact that for $\sin(\phi(x))\ll 1$,
Eq. (\ref{acpsi}) contains $\phi(x)$ to the first order while Eqs.
(\ref{ac1}) contain no $\phi(x)$ dependence, etc. }

\subsection{Identifying the non-locality with gravity --- a conjecture}

We are now faced with the problem of identifying the
phase field $\phi(x)$ with a physical field. Towards this end we 
shall put forward our 
own arguments, and find that supportive formal arguments already exist in 
literature.\cite{Sard,Deh}

As a final movement forward in our thesis, we take note again that
the requirement of parity conservation collapses the phase
$\zeta(x)$ to $\pm 1$ to obtain a Dirac field throughout the
space-time. It is tempting to suggest the phase $\pm 1$ to be
somehow related to the two outstanding facts: (a) The metric of
space-time known in the absence of gravitation is
$(+1,-1,-1,-1)$,\fnm{f}\fnt{f}{It is of no physical relevance to
take the space-time metric as $\mbox{diag.}(-1,+1,+1,+1)$. Parallel
with this observation stands the fact only the {\it relative}
instrinsic parties are of physical signifiacnce.} (b) Only in one
time, and three space dimensions, one finds equal numbers of
generators of rotation and boost. This is due to this circumstance
that the universal covering of the Lorentz group is essentially
given by the chiral group $SU_R(2)\otimes SU_L(2)$ (see
Ref.[\refcite{LHR1987}] for further discussion). For that reason
introducing relative intrinsic parities within the
particle-antiparticle pair gets possible. In flat space-time the
relative intrinsic parties of the particle-antiparticle pair and
the signature of the space-time are deeply intertwined. We,
therefore, suspect that the space-time dependence of $\zeta(x)$ is
closely related to the metric of space-time in the presence of  the
gravitating sources.

Further, in Ref. [\refcite{mod1}] we argued that in the quantum
realm gravitation must introduce an ``{\it in-principle}
unavoidable'' non-local element. Specifically, this non-local
element should appear via modification of the commutativity, or
anti-commutativity, of the fields. Similar conclusions have been
arrived at in the context of string theories, leading to
modification of the fundamental uncertainty relations
as briefly discussed in Sec. 1.

Given  these observations, we tentatively put forward the
conjecture of identifying the discovered non-locality in the
$(1/2,0)\oplus(0,1/2)$ representation space with gravity. Our
conjecture consists of the proposal that the indicated non-locality
should not be used to fix the phase $\phi(x)$ to the integral
multiple of $ \pi$, or $2\pi$ for the conventional Dirac limit, in
order to recover locality.\fnm{g}\fnt{g}{{\it cf.} Secs. 12.5 and
23.6 of Ref. [\refcite{SWQFT}].} Instead, we propose that in the
weak gravitational environments one has to approximate the relative
phase by \begin{equation}
\xi(x)\approx\pm \left[1\pm
i\phi(x)\right]
, \label{grav}
\end{equation}
and identify $\phi(x)$ with $2 G M/c^2 r$ (up to a factor of the
order of unity perhaps). Here, $r$ refers to coordinate distance of the
region of experimental environment from a gravitational source of
mass $M$.\fnm{h}\fnt{h}{Here we have taken liberty of making the
speed of light explicit, and we do not worry about whether the
source is spherical or not, and whether it rotates, etc. Our
interest is essentially qualitative in that regard.} Recalling that
Majorana particles carry {\it imaginary} intrinsic parity, while
the Dirac particles possess {\it real} relative intrinsic parity,
the appearance of $i$ in $\xi(x)\approx\pm
\left[1\pm i\phi(x)\right]$ is interpreted as a direct indication
of the deviation of the particle's intrinsic C and P properties
from the purely Dirac type, and towards the Majorana type (in the
limit $\phi(x)\rightarrow 0$, $\xi(x)$ is immediately seen
to be the relative intrinsic parity).

The gravity-induced CP-violating effects vary from gravitational
environment to gravitational environment. So, while these
CP-violating effects are expected to be large in the vicinity of
neutron stars and the early universe, they are tiny in the
terrestrial environment. The exact magnitude of these CP-violating
effects shall depend on the specific context, but it is expected to
depend on the combination
 $\left(m c^2/E\right)\sin\left( 2GM/c^2 r\right)$.
Whether or not this CP violation is energy-independent, shall
depend upon whether or not a derivative coupling  is considered. In
this context it is to be noted that in experiments (E-82 and E-425)
where the Kaon beam was not horizontal, but entered the ground at
an angle to the horizontal, there remains an ``anomalous
energy-dependence of the Kaon parameters;'' a dependence that can
be further checked by new and carefully planned experiments as
argued by Fischbach and Talmadge.\cite{FT}

The $\beta $ decay processes
$n\rightarrow p+e^-+\overline{\nu}_e$ and $\overline{n}\rightarrow
\overline{p}+e^+ +\nu_e$ which appear conjugated under the CP
transformation, can allow in the presence of gravity-induced CP
violation, processes like:
$\overline{n}\rightarrow p+e^-+\overline{\nu}_e$ and $n\rightarrow
\overline{p}+e^+ +\nu_e$,
where the baryon number is no longer conserved.
Similarly, in the corresponding inverse
$\beta$ decays, gravity-induced CP violation would lead to
lepton-number violating nuclear reactions. It is to be noted that
gravitationally induced neutrino oscillations already respect
lepton flavor oscillations (see [\refcite{grf96}], and last Ref. of
[\refcite{essential}]). In addition, the indicated CP-violating
nuclear reactions generate anti-matter in the matter-rich
environment. Depending on the exact size of these gravity-induced
CP violations, this last observation could provide an efficient
mechanism for converting a part of the neutron star into  gamma
rays and neutrinos in a manner recently suggested by Pen, Loeb, and
Turok.\cite{PLT}

Now, once baryon number non--conservation is addressed, one may
consider a linear superposition of a fermion and an antifermion in
analogy to the Kaon system. {\it Referring to Eqs. (\ref{grav}) and
(\ref{chiral}) one immediately infers that the gravitational phase
carried by the fermion is opposite to that of the antifermion
without invoking antigravity.} This suggests that the apparent
success of the Chardin-Fischbach antigravity framework, in
explaining the observed CP violation for the Kaon system, may lie
(once one goes to the underlying quark level) in the italacised
observation above.

If our framework is realized in nature, then the gravity-induced CP
violation  provides the dynamical reason for how a baryon and
lepton number carrying neutron star  collapses into a black hole
and looses information on the baryonic and leptonic characteristics.
The gravity-induced CP violating nuclear reactions may have
important consequences for the collapse of a black hole into a
space-time singularity. They may indeed prevent the formation of
such a singularity. In addition, the cosmologically observed
matter-antimatter asymmetry would also owe its origins to the
gravity-induced CP violation.

\subsection
{Argument Regarding the Observability of Constant Gravitational
Potentials}

The above conjecture requires us to comment on what contributions
are to be considered in $\phi(x)$. The answer is {\em all}
possible contributions. The local
galactic cluster, known as the Great attractor, embeds us in a
dimensionless gravitational potential $\phi_{GA}/c^2\sim -3 \times
10^{-5}$, see Ref. [\refcite{repeat4}]. Whereas, the same quantity 
for Earth is $\sim -7\times 10^{-10}$, and it is 
$\sim -2\times 10^{-6}$ for the Sun. For experiments performed in the 
vicinity of Earth, or the Solar system, one finds 
\begin{eqnarray}\vec\nabla\phi_{GA}  &\ll& \vec
\nabla\phi_{\{Sun,\,\, Earth\}},\\
 \phi_{GA} &\gg&
\phi_{\{Sun,\,\,Earth\}},
\end{eqnarray} with
$\phi_{\{Sun,\,\,Earth\}}$ standing for the gravitational potentials of
the Sun, 
or Earth, at the experimental site. As a result, for most
{\em classical} experiments (such as orbits of Moon, or planets) the
essentially constant gravitational potential $\phi_{GA}$ has no physical
consequence. Other examples of constant gravitational potentials
that arise in general relativity are known under  the name
of homoids.

However, in 1990 Kenyon\cite{repeat4} emphasised the observability of constant
gravitational potentials in the context of gravitationally induced 
CP violation. This suggestion was strongly questioned by
Nieto and Goldman in their classic 1991 {\em Physics Report}.\cite{NG} In
particular, Nieto and Goldman, following the canonical wisdom,
objected that no  independent experimental means are available to measure
absolute gravitational potentials. These authors, however, apparently 
failed to realize that weak field limits of classical gravity and {\em
any} theory of quantum gravity have different behaviour with respect
to the gravitational potential. While the classical weak-field
limit contains gradient of the gravitational potential, the quantum
weak-field limit contains the gravitational potential itself.
On these grounds it was shown in Ref. [\refcite{EPRg}] that 
the general relativistic description of gravitation turns out
to be incomplete. This incompleteness, it was further argued in the 
previous work, allows for independent experiments to measure  the  constant
gravitational potentials.

Further, it is one of the fundamental assumptions
of Einstein's theory of gravitation that the freely falling
frames are independent of a frame's location. This independence has
been  termed {\em local position invariance}, LPI, by Mann.\cite{Mann} The
incompleteness argument of Ref. [\refcite{EPRg}] shows that LPI is 
violated by the existence of homoid potentials. As an example, 
this happens because a freely falling frame outside the homoid
cavity (inside of which the gravitational potenial is constant, and
varies outside with radial distance from center of the cavity) 
is not identical to a freely falling frame inside the cavity.
The violation of LPI is also manifest
in the weak-field limit that  all quantum theories of gravitation must
satisfy,
and it
leads to physical observability of the homoid, or homoid-like,
gravitational potentials.  Thus the Nieto-Goldman objection to
the Kenyon argument is overcome by the LPI violaton contained in the
incompleteness of Einstein's theory of gravitation.



\section{Conclusion}
\nobreak
\noindent
It is explicitly confirmed, following a remark by Wigner\cite{EPW1963} 
that a representation space carries more
information than a wave equation. Both the celebrated Dirac
equation and the CP-violating Eq. (\ref{d}) belong to the
$(1/2,0)\oplus(0,1/2)$ representation space, and yet both carry
different C, P, and T properties. Similarly, the extended Majorana
spinors and the Dirac spinors belong to the $(1/2,0)\oplus(0,1/2)$
representation space and yet each set has  different physical
properties and describe different physics. 
The core of the physics is contained in the choice of the C, P, and T
properties of the spanning basis spinors. This situation is paralelled
in the $(u,d,c,s)$-flavor space of the quarks as
is immediately seen from a recent work of Kirchbach.\cite{MK}
There the difference
in the choice for the flavor symmetry generators (Gell-Mann's versus
Weyl's) is observable and is 
revealed through the $\eta$-$N$ coupling constant.
The underlying reason for this is that while the
dimensionality of an irreducible representation space does not depend
upon the concrete realization of the symmetry generators, the Noether currents
(Dirac, versus Majorana, versus the CP violating construct,
in space-time; and  Gell-Mann versus Weyl in the flavor space) do.
Taking the $(1/2,0)\oplus(0,1/2)$ representation space as a study case, we see
that in going from pure right- and left-handed $(1/2,0)$ and
$(0,1/2)$ spinors to their direct sum $(1/2,0)\oplus(0,1/2)$, one
transports information about the C, P, and T properties of the
representation space as well as about the properties of the theory
with respect to locality. Having derived the CP-violating wave
equation for spin one-half, one could have demanded locality, and
could have recovered Dirac equation. However, we have taken a
different path to proceed, and conjecture the non-locality in
quantum field theory to originate in a specific manner from
gravity. From that one predicts that there exists a CP violation
that varies from gravitational environment to gravitational
environment, remaining small in the terrestrial environment, and
becoming significantly large in the vicinity of neutron stars and
the early universe.

In summary, we have presented a thesis  on an origin of CP
violation that lies at the  level of the representations of the
Lorentz group, and is related to the space-time metric in the
presence of a gravitational source. A CP violation that depends on
the gravitational environment via the factor $ GM/c^2 r$
shall have dramatic astrophysical and cosmological implications.
Especially, because this factor can vary from about $0.2$
for the surface of a $1.4$ solar mass neutron star
down to roughly $ 10^{-9}$ for Earth's surface. The last 
number can, however, increase to roughly $10^{-5}$ 
if the contribution from the great attractor
is taken into account (as it must be).
Gravity induced CP violation shall alter the equation of state for
nuclear matter in intense gravitational fields and hence the fate
of neutron stars, supernovae, and may be an important physics factor
in the cosmic gamma ray bursts.

\nonumsection{Acknowledgements}

\noindent
I am grateful to Gabriel Chardin, 
Recai Erdem, and Ephraim Fischbach
for a reading of the manuscript and for providing to me 
their suggestions. 
This work was done, in part, under the auspices of the U.S. Department of 
Energy and supported, in part, by the Mexican CONACyT.

\nonumsection{References}

\end{document}